\documentclass[final,1p,times]{elsarticle} 
\usepackage{graphicx} 
\usepackage{amssymb} 
\usepackage{amsthm} 
\usepackage{lineno} 

\journal{Nuclear Physics A} 
\begin{document} 
\begin{frontmatter} 
\title{$J/\psi$ production at mid and forward rapidity at RHIC}
\author{Zhen Qu$^1$}
\author{Yunpeng Liu$^1$}
\author{Nu Xu$^2$}
\author{Pengfei Zhuang$^1$}
\address{
$^1$Physics Department, Tsinghua University, Beijing
100084, China\\ $^2$Nuclear Science Division, Lawrence Berkeley
National Laboratory, Berkeley, California 94720, USA}
\begin{abstract} 
We calculate the rapidity dependence of $J/\psi$ nuclear
modification factor and averaged transverse momentum square in heavy
ion collisions at RHIC in a 3-dimensional transport approach with
regeneration mechanism.
\end{abstract} 
\date{\today}
\end{frontmatter} 

$J/\psi$ suppression~\cite{matsui} is widely accepted as a probe of
quark-gluon plasma (QGP) formed in relativistic heavy ion collisions
and was first observed at SPS~\cite{gonin} more than ten years ago.
At RHIC and LHC energy, a significant number of charm quarks are
generated in central heavy ion collisions, and the recombination of
these uncorrelated charm quarks offers another source for $J/\psi$
production~\cite{thews}. There are different ways to describe the
$J/\psi$ regeneration. In the statistical model~\cite{pbm}, all the
$J/\psi$s are produced at hadronization of the system through
thermal distributions and charm conservation. In some other models,
$J/\psi$s come from both the continuous regeneration inside the hot
medium and primordial production through initial nucleon-nucleon
collisions~\cite{thews, grandchamp}. The regeneration is used to
describe~\cite{yan} the $J/\psi$ nuclear modification factor
$R_{AA}$ and averaged transverse momentum square $\langle
p_t^2\rangle$. From the experimental data~\cite{phenix1} at RHIC,
the $R_{AA}$ is almost the same as that at SPS, which seems
difficult to explain in models with only primordial production
mechanism.

The rapidity dependence of $J/\psi$ production was also measured at
RHIC~\cite{phenix1,phenix2} and discussed in
models~\cite{kharzeev,andronic,zhao}. The surprising finding of the
experimental result is that the apparent suppression at forward
rapidity is stronger than that at midrapidity, i.e. $R_{AA}^{mid} >
R_{AA}^{forward}$. This is again difficult to explain in models with
only initial production mechanism, since the suppression at
midrapidity should be stronger than that at forward rapidity. Not
only $R_{AA}$ but also $\langle p_t^2\rangle$ depends on the
rapidity~\cite{phenix1}. In semi-central and central Au+Au
collisions the value of $\langle p_t^2\rangle$ at midrapidity is
lower than that at forward rapidity, i.e. $\langle
p_T^2\rangle^{mid} < \langle p_t^2\rangle^{forward}$. In this paper,
we will discuss the rapidity dependence of $R_{AA}$ and $\langle
p_t^2\rangle$ in a 3-dimensional transport model with both initial
production and continuous regeneration mechanisms.

At RHIC $J/\psi$s are detected at midrapidity $|y|<0.35$ and
forward rapidity $1.2<y<2.2$, both are located in the plateau of
the rapidity distribution of light hadrons~\cite{brahms}.
Therefore, the space-time evolution of the QGP can be
approximately described by the transverse hydrodynamic equations
at midrapidity~\cite{zhu}, and the $J/\psi$ motion is controlled
by a 3-dimensional transport equation in an explicitly boost
invariant form
\begin{equation}
\left[\cosh(y_\Psi-\eta)\
\partial/\partial\tau+\sinh(y_\Psi-\eta)/\tau\ \partial/\partial
\eta+{\bf v}_t^\Psi\cdot\nabla_t\right]f_\Psi =-\alpha_\Psi
f_\Psi+\beta_\Psi,
\end{equation}
where $f_{\Psi}=f_\Psi({\bf p}_t,y,{\bf x}_t,\eta,\tau|{\bf b})$
is the charmonium distribution function in phase space at fixed
impact parameter ${\bf b}$, and we have used transverse energy
$E_t=\sqrt{E_\Psi^2-p_z^2}$, rapidity
$y_{\Psi}=1/2\ln[(E_\psi+p_z)/(E_\Psi-p_z)]$, proper time
$\tau=\sqrt{t^2-z^2}$ and space-time rapidity
$\eta=1/2\ln{[(t+z)}/{(t-z)]}$ to replace the charmonium energy
$E_\Psi=\sqrt{{\bf p}^2+m_\Psi^2}$, longitudinal momentum $p_t$,
time $t$ and longitudinal coordinate $z$. The term with transverse
velocity ${\bf v}_t^\Psi={\bf p}_t/E_t$ reflects the leakage
effect in charmonium motion. To take into account the decay of the
charmonium excitation states into $J/\psi$, the symbol $\Psi$ here
stands for $J/\psi, \chi_c$ and $\psi'$ and the ratio of their
contributions in the initial condition is taken as 6:3:1. The
suppression and regeneration in the QGP are described by the loss
and gain terms on the right hand side of the transport equation.
Considering the gluon dissociation process, $\alpha$ can be
explicitly written as~\cite{liu}
\begin{equation}
\alpha_{\Psi}({\bf p},{\bf x}, t|{\bf b}) = \int d^3{\bf
q}/\left((2\pi)^3 4E_tE_g\right)W_{g\Psi}^{c\bar c}(s)f_g({\bf
q},T,u)\Theta(T-T_c)/\Theta(T_d^\Psi-T),
\end{equation}
where $W_{g\Psi}^{c\bar{c}}$ is the transition
probability~\cite{peskin} as a function of the colliding energy
$\sqrt s$ of the dissociation process, $E_g$ and $f_g$ are the
gluon energy and gluon thermal distribution, and $T_c$ and
$T_d^\Psi$ are the critical temperature of the deconfinement phase
transition and dissociation temperature of $\Psi$, taken as
$T_c=165$ MeV,\ \ $T_d^{J/\psi}/T_c=1.9$ and
$T_d^{\chi_c}/T_c=T_d^{\psi'}/T_c=1$. The two step functions
$\Theta$ in $\alpha$ indicate that the suppression is finite in
the QGP phase at temperature $T<T_d^\Psi$ and becomes infinite at
$T>T_d^\Psi$. We have here neglected the suppression process in
hadron phase~\cite{zhu,yan}.

The gain term $\beta_{\Psi}$ can be obtained from the loss term
$\alpha$ by considering detailed balance~\cite{thews}. We assume
local thermalization of charm quarks in the QGP and take the charm
quark distribution as
\begin{equation}
f_c({\bf k},{\bf x},t)={\rho_c({\bf x},t)}f_q({\bf k})\label{eq:4}
\end{equation}
with $\rho_c$ being the density of charm quarks in coordinate space,
\begin{equation}
 \rho_{c}({\bf x},t)=T_A({\bf x}_t)T_B({\bf x}_t-{\bf b})\cosh\eta/\tau\
d\sigma_{NN}^{c\bar{c}}/d\eta
\end{equation}
and $f_q$\ the normalized Fermi distribution in momentum space,
where $T_A$ and $T_B$ are the thickness functions for the two
colliding nuclei determined by nuclear geometry. Since the large
uncertainty of charm quark production cross section in pp collisions
for both experimental and theoretical studies, we assume the
rapidity dependence of charm production as a Gauss distribution
$d\sigma_{pp}^{c\bar{c}}/d\eta=d\sigma_{pp}^{c\bar{c}}/d\eta\big{|}_{\eta=0}
e^{-\eta^2/2\eta_0^2}$ with
$d\sigma_{pp}^{c\bar{c}}/d\eta\big{|}_{\eta=0}=120\ \mu$b which
agrees with the experimental data~\cite{zhang} and
$(d\sigma_{pp}^{c\bar{c}}/d\eta\big{|}_{\eta=1.7})/(d\sigma_{pp}^{c\bar{c}}/d\eta\big{|}_{\eta=0})
=1/3$ to determine the parameter $\eta_0$ which is in between the
smallest and largest theoretical estimation~\cite{zhang}.

The contribution from the primordial charmonium production is
reflected in the initial condition of the transport equation at
the starting time $\tau_0$.  By fitting the experimental
data~\cite{phenix3} for pp collisions at RHIC, the initial
charmonium momentum distribution is extracted as
\begin{equation}
 f_{pp}({\bf p}_t,y)=5g(y)/\left(4\pi \langle p_t^2\rangle(y)\right)\left[1+p_t^2/\left(4\langle
 p_t^2\rangle(y)\right)\right]^{-6},
\end{equation}
where the rapidity distribution $g(y)$ is a double Gauss
function~\cite{phenix3}, and the rapidity dependence of the
averaged transverse momentum square is taken as $\langle
p_t^2\rangle(y)=\langle p_t^2\rangle(0)(1-y^2/y_{max}^2)$ with the
parameters $\langle p_t^2\rangle(0)=$ 4.1 (GeV/c)$^2$ and
$y_{max}=\textrm{arccosh}(\sqrt{s}/{2m_{J/\psi}})$. Note that, to
include the Cronin effect in the initial state of heavy ion
collisions~\cite{gavin}, we add an extra term to $\langle
p_t^2\rangle$ which comes from the gluon multi-scattering with
nucleons~\cite{yan,zhao}.

The charmonium production, including initial production and
regeneration, is related to the QGP evolution through the local
temperature $T$ and fluid velocity $u_\mu$ appearing in the thermal
gluon and charm quark distributions, they are determined by the
ideal hydrodynamics~\cite{zhu}.

\begin{figure}[!hbt]
\centering
\includegraphics[width=0.44\textwidth]{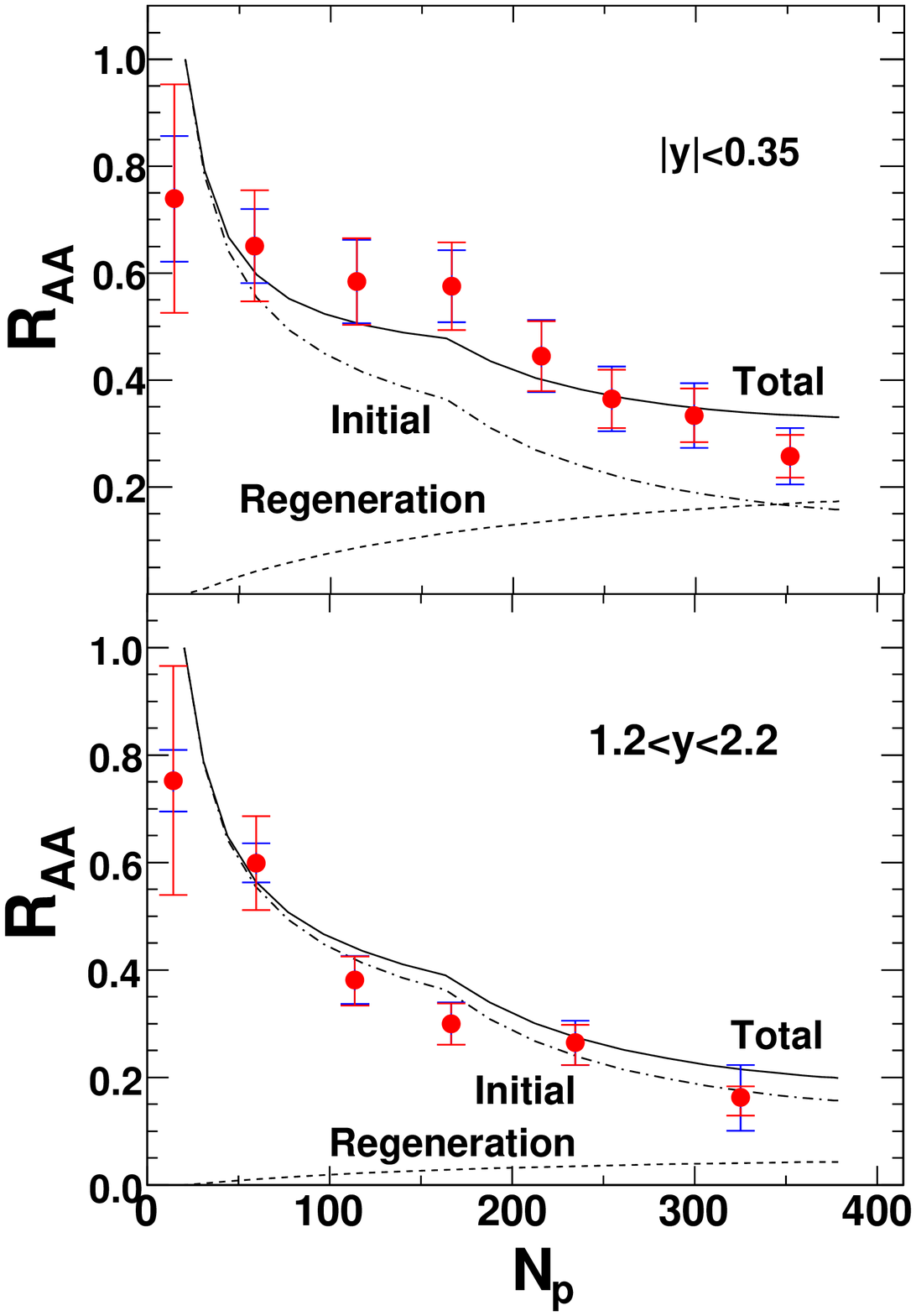}%
\hspace{0mm}%
\includegraphics[width=0.485\textwidth]{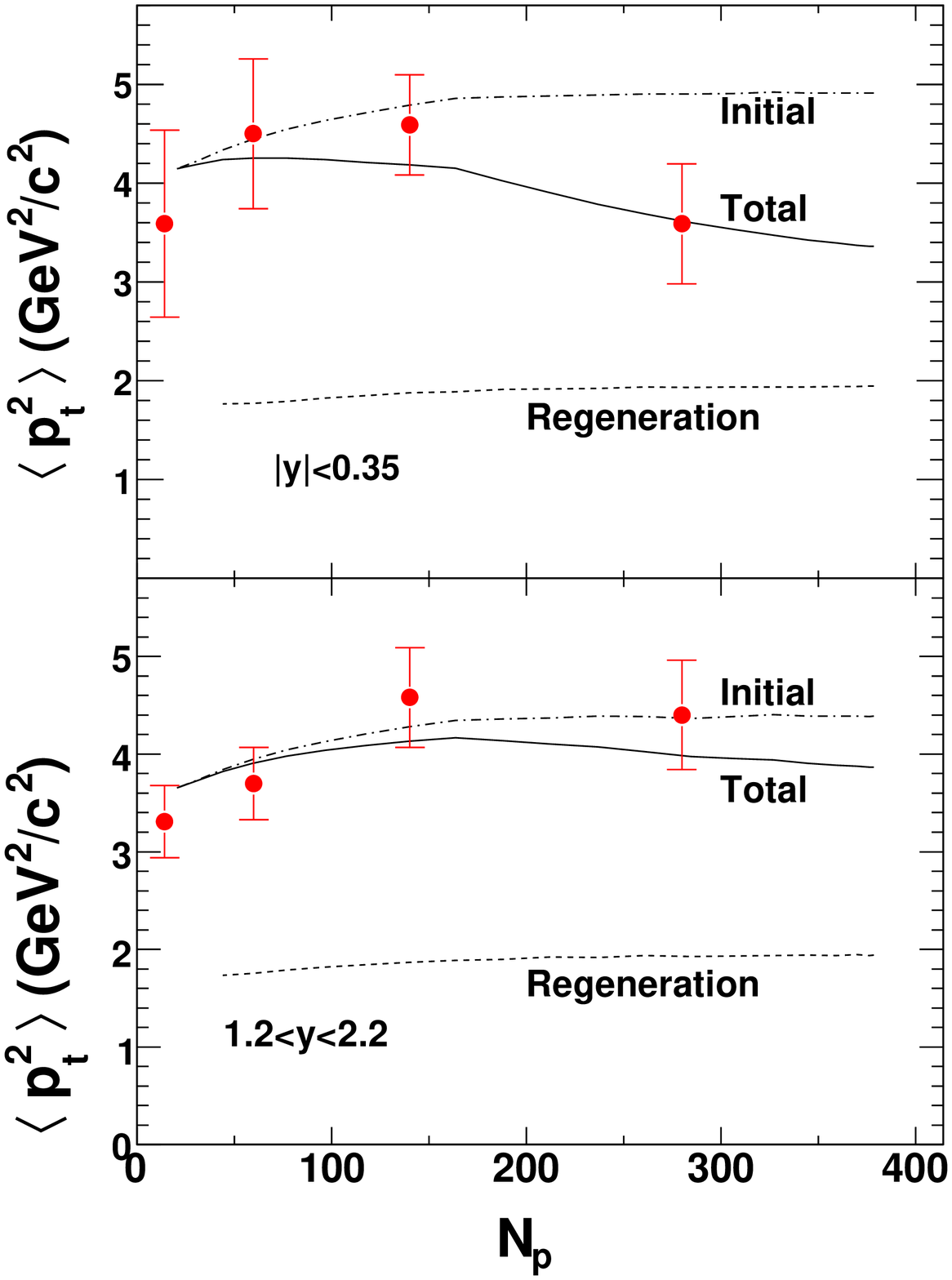}
\caption{The nuclear modification factor $R_{AA}$ (left panel) and
averaged transverse momentum square $\langle p_t^2 \rangle$ (right
panel) at mid and forward rapidity as functions of number of
participants $N_p$. The theoretical calculations with only initial
production (dot-dashed lines), only regeneration (dashed lines)
and both (solid lines) are compared with the experimental
data~\cite{phenix1,phenix2}.}\label{fig1}
\end{figure}

With the known distribution $f_{J/\psi}({\bf p}_t,y,{\bf
x}_t,\eta,\tau|{\bf b})$, one can calculate the $J/\psi$ yield and
momentum spectra. The nuclear modification factor $R_{AA}$ and
averaged transverse momentum square $\langle p_t^2\rangle$ at mid
and forward rapidity are shown in Fig.\ref{fig1} as functions of
centrality. Since $R_{AA}$ is normalized to the pp collisions, the
assumption of the same medium at mid and forward rapidity leads to
similar $R_{AA}$ in the two rapidity regions, when we consider
only initial production, as shown in the left panel. The
regeneration at forward rapidity is, however, much less than that
at mid rapidity. As a result of the competition, the total
$R_{AA}$ at forward rapidity is less than that at mid rapidity,
consistent with the experimental observation.

While the population is dominated by low momentum $J/\psi$s, the
averaged transverse momentum carries more information on high
momentum $J/\psi$s and can tell us more about the dynamics of
charmonium production and suppression. The initially produced
$J/\psi$s are from the hard nucleon-nucleon process at the very
beginning of the collision and their $p_t$ spectrum is harder.
From the gluon multi-scattering with nucleons before the two
gluons fuse into a $J/\psi$, there is a $p_t$ broadening for the
initially produced $J/\psi$s. Considering further the leakage
effect which enables the high momentum $J/\psi$s escape from the
anomalous suppression in the hot medium, the initially produced
$\langle p_t^2\rangle$ increases smoothly with centrality and
becomes saturated at large $N_p$. Since the regenerated $J/\psi$s
are from the thermalized charm quarks inside the QGP, their
averaged momentum is small and almost independent of the
centrality. Both the initially produced and regenerated $\langle
p_t^2\rangle$ is not sensitive to the rapidity region. While the
difference between the initially produced and regenerated $R_{AA}$
decreases with increasing $N_p$, the difference between the values
of $\langle p_t^2\rangle$ from the two rapidity regions increases
smoothly with centrality! The total $\langle p_t^2\rangle$ depends
strongly on the fraction of the regeneration. At mid rapidity, the
regeneration and initial production are equally important in
central collisions, see the left panel of Fig.\ref{fig1}. The
large contribution from the regeneration leads to a remarkable
decrease of the value of $\langle p_t^2\rangle$ at mid rapidity.
At forward rapidity, the regeneration contribution is, however,
very small even for central collisions, see the left panel again.
In this case, the total $\langle p_t^2\rangle$ is dominated by the
initial production in the whole $N_p$ region.

In summary, we calculated the $J/\psi$ nuclear modification factor
and averaged transverse momentum square at mid and forward
rapidity in a three dimensional transport approach. The
experimentally observed rapidity dependence of $R_{AA}$ and
$\langle p_t^2\rangle$ in Au+Au collisions at $\sqrt s$=200 GeV
can well be explained by our model calculation where the
continuous regeneration of $J/\psi$ from thermalized charm quarks
in QGP is an important ingredient. We predict that at higher
colliding energies, for example at LHC, the regeneration will
become the dominant ingredient.


\section*{Acknowledgments}
We are grateful to Dr. Xianglei Zhu for the help in numerical
calculations. The work is supported by the NSFC grant No.
10735040, the National Research Program Grants 2006CB921404 and
2007CB815000. and the U.S. Department of Energy under Contract No.
DE-AC03-76SF00098.


\end{document}